\documentclass[journal]{IEEEtran}

\usepackage{cite}
\usepackage[pdftex]{graphicx}
\usepackage[cmex10]{amsmath}
\usepackage{algorithmicx}
\usepackage{array}
\usepackage[caption=false,font=normalsize,labelfont=sf,textfont=sf]{subfig}
\usepackage{fixltx2e}
\usepackage[T1]{fontenc}
\usepackage[latin1]{inputenc}                       
\usepackage{lscape}
\usepackage{latexsym,exscale,stmaryrd,amssymb,amsfonts,amstext,amsthm}
\usepackage{stmaryrd}	
\usepackage{dsfont}
\usepackage{algpseudocode}
\usepackage{mathtools}
\usepackage{paralist} 
\usepackage[english]{babel}
\usepackage{tikz}
\usetikzlibrary{matrix}
\usetikzlibrary{arrows}
\graphicspath{ {Plots/} }

\algnewcommand\algorithmicinput{\textbf{Input:}}
\algnewcommand\Input{\item[\algorithmicinput]}

\newcommand{\abs}[1]{\ensuremath{\left\vert#1\right\vert}}
\newcommand\ZuWeis{\mathrel{\mathop:\!\!=}} 
\newcommand\WeisZu{\mathrel{=\!\!\mathop:}} 
\newcommand{\N}{\ensuremath{\mathds{N}}}  
\newcommand{\R}{\ensuremath{\mathds{R}}} 
  
\newcommand{\Pp}{\ensuremath{\textbf{P}}}  
\newcommand{\rank}{\operatorname{rank}} 
\newcommand{\spa}{\operatorname{span}} 
\newcommand{\norm}[1]{\|#1\|}

\newcommand{\Mod}[1]{\operatorname{mod}\; #1}

\newcommand{\fG}{\ensuremath{\mathfrak G}}

\newcommand{\fo}{\ensuremath{\mathfrak o}}

\newcommand{\fR}{\ensuremath{\mathfrak R}}
\newcommand{\fr}{\ensuremath{\mathfrak r}}

\newcommand{\fs}{\ensuremath{\mathfrak s}}
\newcommand{\fT}{\ensuremath{\mathfrak T}}

\newcommand{\Al}{{\ensuremath{\Omega }}}
\newcommand{\al}{{\ensuremath{ \omega}}}
\newcommand{\M}{{\ensuremath{ P}}}
\newcommand{\iM}{{\ensuremath{ p}}} 
\newcommand{\m}{{\ensuremath{ N}}} 
\newcommand{\im}{{\ensuremath{n}}} 
\newcommand{\imp}{{\ensuremath{ m}}} 
\newcommand{\n}{{\ensuremath{ T}}} 
\newcommand{\iN}{{\ensuremath{ t}}} 
\newcommand{\Wei}{{\ensuremath{ A}}} 
\newcommand{\wei}{{\ensuremath{ a}}}
\newcommand{\Sou}{{\ensuremath{ S}}} 
\newcommand{\Mix}{{\ensuremath{ X}}} 
\newcommand{\mix}{{\ensuremath{ x}}} 
\newcommand{\sAl}{{\ensuremath{ \fs}}} 
\newcommand{\sAlv}{{\ensuremath{ \fr}}} 
\newcommand{\Nn}{{\ensuremath{ M}}}

\newcounter{mod}

\newcounter{aaa}

\newtheorem{lem1}{Lemma} \numberwithin{lem1}{section} 
\newtheorem{cor1}[lem1]{Corollary}
\newtheorem{theo1}[lem1]{Theorem}
\newtheorem{example1}[lem1]{Example}

\newtheorem{df1}[lem1]{Definition}

\newtheorem{remark1}[lem1]{Remark}

\newenvironment{assum}{\refstepcounter{aaa}\par\textbf{A \theaaa.}}{}

\newenvironment{example}[1]{\begin{example1}[#1]\ \newline}{\end{example1}}
\newenvironment{exampleon}{\begin{example1}\ \newline}{\end{example1}}  
\newenvironment{remark}{\begin{remark1}\ \newline}{\end{remark1}}
\newenvironment{theo}{\begin{theo1}\ \newline}{\end{theo1}}

\newenvironment{lemon}{\begin{lem1}\ \newline}{\end{lem1}}

\newenvironment{dfon}{\begin{df1}\ \newline}{\end{df1}}

\begin{document}

\title{Identifiability for Blind Source Separation of Multiple Finite Alphabet Linear Mixtures}

\author{Merle~Behr,
        Axel~Munk 
\thanks{M. Behr is with the Institute for Mathematical Stochastics and the Felix Bernstein Institute for Mathematical Statistics in the Bioscience, University of Goettingen, 37077 G\"ottingen, Germany (e-mail: behr@math.uni-goettingen.de).
A. Munk is with the Institute for Mathematical Stochastics, the Felix Bernstein Institute for Mathematical Statistics in the Bioscience, University of Goettingen and the Max Planck Institute for Biophysical Chemistry, 37077 G\"ottingen, Germany (e-mail: 	munk@math.uni-goettingen.de). }}


\maketitle

\begin{abstract}
We give under weak assumptions a complete combinatorial characterization of identifiability for linear mixtures of finite alphabet sources, with unknown mixing weights and unknown source signals, but known alphabet. 
This is based on a detailed treatment of the case of a single linear mixture. Notably, our identifiability analysis applies also to the case of unknown number of sources.
We provide sufficient and necessary conditions for identifiability and give a simple sufficient criterion together with an explicit construction to determine the weights and the source signals for deterministic data by taking advantage of the hierarchical structure within the possible mixture values. 
We show that the probability of identifiability is related to the distribution of a hitting time and converges exponentially fast to one when the underlying sources come from a discrete Markov process. 
Finally, we explore our theoretical results in a simulation study.
Our work extends and clarifies the scope of scenarios for which blind source separation becomes meaningful.
\end{abstract}

\begin{IEEEkeywords}
Blind source separation; BSS; finite alphabet signals; single mixture; instantaneous mixtures; Markov processes, stopping time 
\end{IEEEkeywords}

\section{Introduction}

\subsection{Problem description}

In this work we are concerned with identifiability in a particular kind of blind source separation (BSS) motivated by different applications in digital communication (see e.g.,\cite{verdu, proakis, zhang}), but also in cancer genetics (see e.g.,\cite{yau, carter, liu, titan }). 
A prominent example is the separation of a mixture of audio or speech signals, which has been picked up by several microphones, simultaneously (see e.g., \cite{aissa}). In this case the different speech signals correspond to the sources and the recordings of the microphones to the mixture of signals with unknown mixture weights. From this mixture the individual signals have to be separated.

More generally, in BSS problems one observes $\M$ mixtures of $\m$ sources and aims to recover the original sources from the available observations. In this paper we focus on the linear case  (for the non-linear case see e.g., \cite{lachover, castella}), where the blindness refers to the fact that neither the sources nor the mixing weights are known.  We also treat the case of unknown number of sources.


A minimal requirement underlying any recovery algorithm for sources and mixture weights (in a deterministic or noisy setting) to be valid is identifiability, i.e., the unique decomposition of the mixture into sources and mixing weights.
Without any additional information on the source signals $\Sou$ identifiability cannot hold, of course, as from a linear system $\Mix = \Wei \Sou$ the matrices $\Wei$ and $\Sou$ are not uniquely determined, in general. In particular this applies to single linear mixtures ($\M = 1$). However, if we assume that the values of the sources are attained in a known finite set of real numbers (finite alphabet), then BSS identifiability holds under certain conditions, which is necessary for any recovery algorithm to be valid. 

Therefore, the aim of this work is to give a comprehensive discussion of such conditions. We are able to give (under weak assumptions) a complete combinatorial characterization of identifiability. From this we derive sufficient conditions, which are easy to verify. 
Moreover, these conditions yield an explicit construction to recover sources and weights in the noiseless case from the deterministic mixture.
More specifically, using the notation
\begin{align*}
v = (v_i)_{1\leq i \leq n} = (v_1,\ldots,v_n)^{\top}, \quad V = (V_{ij})_{\substack{1\leq i \leq n \\ 1\leq j \leq m}} 
\end{align*}
for a vector $v \in \R^n$ and a matrix $V \in \R^{n\times m}$ with $n$ rows and $m$ columns and $V_{i \cdot}, V_{\cdot j}$ for the corresponding row and column vectors, respectively,
we assume from now on that the observed signal is linked to the sources via
\begin{align}\label{model}
\Mix= \Wei \Sou 
\end{align}
where $\Mix$ is the deterministic mixture, $\Wei =(\Wei_{\iM  \im})_{1\leq \iM \leq \M, \; 1\leq \im \leq \m}$ are the unknown mixing weights, $\Sou=(\Sou_{1 \cdot},\ldots,\Sou_{\m \cdot})^\top \in \Al^{\m \times \n}$ are the unknown source signals, and $\Al \ZuWeis \{\al_1,\ldots,\al_k\}\subset \R$ is a known alphabet, i.e., the set of possible values the sources can attain. 
The row vectors $\Sou_{\im \cdot}$ with $1\leq \im \leq \m$ denote the single source signals (of length $\n$) and the row vectors $\Wei_{\iM \cdot}$ with $1 \leq \iM \leq \M$ denote the mixing vectors (of length $\m$) of the different mixtures.

In an observational model we often have $Y = \Mix + \epsilon$ where $\epsilon$ is some random noise term with zero mean. All identifiability results for (\ref{model}) transfer to this situation as well, of course.
However, we stress that the primarily aim of this paper is not to provide a method which reconstructs the mixing weights $\Wei$ and the sources $\Sou$ from the (possibly) noisy observations $Y$, but rather to clarify the scope of scenarios under which this is possible. To this end, we will analyze necessary and sufficient conditions under which the decomposition in (\ref{model}) is unique. 

For the moment, we assume that the number of sources $\m$ is known. However, we point out that most of our results remain true even when $\m$ is unknown (see Section \ref{sec: um}), a case which has never been treated before to best of our knowledge.

\subsection{Related Work}

As far as we are aware of, identifiability of model (\ref{model}) has not yet been considered in the literature in this general form, although various special cases and variations of this BSS problem have been addressed. 
The particular case of a binary alphabet, i.e., when $k=2$ and $\{\al_1,\al_2\}=\{-1,1\}$, has been considered in \cite{telwar, pajunen, diamantras2,diamantras5, gu}.
Diamantaras and Papadimitriou \cite{diamantaras4} and Rostami et al. \cite{rostami} assume the alphabet to be equally spaced, i.e., $\{\al_1,\ldots,\al_k\}=\{\al_0,\al_0+T,\al_0+2T,\ldots,\al_0+kT\}$. We consider arbitrary finite alphabets $\{\al_1,\ldots,\al_k\}\subset \R$.
Moreover, several authors (e.g.,\cite{pajunen, diamantaras4,rostami}) assume a specific distribution on the alphabet, e.g., uniform. Our results show that it is already sufficient to observe some specific combinations of alphabet values (which are minimal conditions in a sense), hence we do not need to assume such a specific distribution.
Diamantaras \cite{diamantras3} works with a general finite alphabet as well but considers a mixture of two sources, $\m=2$. 
Li et al. \cite{amari} gave necessary and sufficient identifiability criteria for sparse signals, i.e., signals having many zero entries, in contrast to our work. Although they consider underdetermined mixtures, their results require at least two sensors ($\M>1$) and do not hold for single linear mixture ($\M=1$). Bofill and Zibulevsky \cite{bofill} suggest for $\M=2$ a method for estimating the mixing weights for sparse signals as well, although without giving any explicit identifiability criteria.
Diamantaras \cite{diamantras5} considers a general finite alphabet but assumes the mixing weights to be complex. Thus, he works with a 2-dimensional signal. Combing the results from \cite{diamantras5} and \cite{diamantaras6} yields a sufficient identifiability criterion for finite alphabet sources with complex mixing weights. However, this result does not hold when the mixing weights are real as considered here.

There are further variations of the BSS problem. Some of them are associated with Independent Component Analysis (ICA) (see e.g., \cite{comon}), which is based on the stochastic independence of the different sources (assumed to be random). ICA can be a powerful tool for (over)determined models ($\M\geq \m$) \cite{comon} and there are approaches for underdetermined  multiple linear mixture models ($1<\m<\M$) as well \cite{lee}. However, ICA is not applicable for single linear mixtures ($\M=1$), as the error terms of the single sources sum up to a single error term such that stochastic independence of the sources becomes irrelevant.

Also conceptually related is blind deconvolution (see e.g., \cite{li2, diamantarasD}), however, the convolution model makes analysis and identifiability severely different \cite{yellin}.

Another related problem is non-negative matrix factorization (see e.g., \cite{lee1999, donoho, arora}), where one assumes (\ref{model}), but instead of $\Sou \in \Al^{\m \times \n}$, both $\Sou$ and $\Wei$ are non-negative. Indeed, the identifiability conditions derived in \cite{donoho,arora} are quite related in nature to ours in Section \ref{subsec:sc}, where their simpliciality condition on $\Wei$ corresponds to our condition (\ref{emptyi}) and their separability condition on $\Sou$ corresponds to our assumption A\ref{assusuff}. However, whereas their assumptions necessarily imply $\M > 1$, ours yield identifiability for single linear mixtures ($\M = 1$), explicitly exploring the finite alphabet.

For the non-blind scenario, i.e., when $\Wei$ in model (\ref{model}) is known, \cite{aissa2} considers identifiability in a probabilistic framework.

To the best of our knowledge, a comprehensive characterization and unifying treatment of identifiability of the mixing weights and the sources for model (\ref{model}) has been elusive. This issue is, however, fundamental for identifying the scope of possible scenarios where recovery algorithms for $\Sou$ and $\Wei$ in (\ref{model}) are applicable  (see, e.g., \cite{ pajunen, diamantras5, gu, rostami}). In this sense our work provides an almost "minimal" set of conditions under which any recovery algorithm for the BSS problem only can be expected to be valid; in the noiseless case as well as for the case with random error.

\subsection{Organisation of the paper}

We will start our analysis by making two simplifications, which lead to a better interpretation of the corresponding identifiability conditions.

First, we will assume that $\M  = 1$, i.e., that we observe a single linear mixture in (\ref{model}). Clearly, when $\M$ increases the identification problem becomes easier, as more mixtures of the same sources are observed. Thus, the case $\M = 1$ corresponds to the most difficult scenario and therefore we treat this case in detail. Generalizations to arbitrary $\M$ will then follow easily from this case and are given in Section \ref{sec:mlm}.

Second, we start with considering probability mixing weights, i.e., $\Wei_{\iM \im}>0$ and $\sum_{\im = 1}^\m  \Wei_{ \iM \im} = 1$ for all $\iM = 1,\ldots,\M$. This is because the corresponding identifiability conditions have an easier interpretation when the mixing weights are positive. When we allow for negative mixing weights, the identifiability issue becomes slightly more difficult and the corresponding conditions become more complicated. Generalizations to negative mixing weights are given in Section \ref{sec:abw}.

The paper is organized as follows.
After introducing a rigorous formulation of the problem and model (Section \ref{sec:probstat}), we will give a necessary and sufficient identifiability criterion for the mixing weights and the sources (Section \ref{subsec:nsc}). To this end, we will first characterize the identifiability issue as a purely combinatorial problem.
In Sections \ref{subsec:sc} we will then generalize a result from Diamantaras and Chassioti \cite{diamantras2} in order to derive a simple sufficient identifiability criterion. This characterizes those source signal combinations which make the variation of the mixture rich enough in order to become identifiable. This condition also provides an explicit construction (see the algorithm in Figure \ref{fig:algodetmu}) for recovery of the weights $\Wei$ and the sources $\Sou$ from the mixture $\Mix$ in (\ref{model}).

In Section \ref{sec:sp} we will shortly discuss how likely it is for the identification criterion of Section \ref{subsec:sc} to be satisfied when the underlying sources $\Sou_{1 \cdot},\ldots,\Sou_{\m \cdot}$ are discrete Markov processes. To this end, we will bound the probability of identification from below by a phase-type distribution. Using a stopping time argument, we will show that the mixture becomes identifiable exponentially fast, which reveals identifiability as less a critical issue in many practical situations as one might expect.

Although, we assume the number of source components $\m$ to be known, in Section \ref{sec: um} we show that most of our results remain true even when $\m$ is unknown.

In Section \ref{sec:sim} we simulate source signals from a two-state Markov chain to illustrate how far our derived sufficient identifiability conditions (Section \ref{subsec:sc}) are from being necessary in this setting. To this end, our results from Section \ref{subsec:nsc} are fundamental as they give an explicit way to decide whether a given mixture is identifiable or not. Our simulation results reveal the simple sufficient condition in Section \ref{subsec:sc} as quite sharp as we find that the number of observations needed for this to hold is quite close to the actual number of observations required for identifiability. This establishes the exponential bound in Section \ref{sec:sp} as a useful tool to estimate the required number of observations guaranteeing identifiability with high probability.
  
We conclude in Section \ref{sec:conc}.

\section{Problem Statement}\label{sec:probstat}
As mentioned above, for ease of presentation, we start with analyzing identifiability in (\ref{model}) for single linear mixtures, i.e., $\M = 1$ (for arbitrary $\M$ see Section \ref{sec:mlm}) and probability mixing weights (for arbitrary mixing weights see Section \ref{sec:abw}).
To this end, let $\mathcal{A} \ZuWeis \{ \wei \in\R^\m : 0<\wei_1<...<\wei_\m < 1 \text{ and } \sum_{\im = 1}^\m \wei_\im = 1 \}$ denote the set of positive (probability) mixing vectors.
Note that $\wei_\im \neq \wei_\imp$ for $\im \neq \imp$ is always necessary to ensure identifiability of different sources $\Sou_{\im \cdot},\Sou_{\imp \cdot}$ (if $\wei_\im = \wei_\imp$ interchanging of $\Sou_{\im \cdot}, \Sou_{\imp \cdot}$ results in the same mixture $\mix$).
For given finite alphabet $\Al=\{\al_1,\ldots,\al_k\}\subset \R$, with $\al_1<\ldots<\al_k$, number of sources $\m \in \N$, and number of observations $\n \in \N$ let the sources $\Sou = (\Sou_{1 \cdot},\ldots,\Sou_{\m \cdot})^\top \in \Al^{\m \times \n} $ and the mixing weights $\wei = (\wei_1,\ldots,\wei_\m) \in \mathcal{A}$. Then the observed values are given by $\mix =  \wei \Sou $, i.e.,
\begin{align}\label{modeld}
\mix_\iN = \sum_{\im = 1}^\m \wei_\im \Sou_{\im\iN}, \qquad \iN = 1,\ldots,\n.
\end{align}
\begin{dfon}
Let $\mix = \wei \Sou $; $(\wei, \Sou) \in \mathcal{A} \times \Al^{\m \times \n} $ as in (\ref{modeld}). Then we denote the vector $\wei$ and the matrix $\Sou$ as (jointly) identifiable from the observational vector $\mix$ when there exists exactly one $(\tilde{\wei},\tilde{\Sou})\in \mathcal{A} \times \Al^{\m \times \n} $ such that $\mix= \tilde{\wei} \tilde{\Sou} $.
\end{dfon}
In other words, identifiability means that $\mix = \tilde{\mix}$ in (\ref{modeld}) implies that $\wei = \tilde{\wei}$ and $\Sou = \tilde{\Sou}$.
For simplicity, we refer to $\wei$ and $\Sou$ being identifiable from $\mix$  just by saying that $(\wei,\Sou)$ is identifiable.
The aim of this paper is to study under which conditions $(\wei, \Sou)$ is identifiable from $\mix$. 

Even though we assumes $\m$ to be known, most of our results remain true when $\m$ is unknown, i.e., $\mix = \wei \Sou$ with $(\wei,\Sou) \in \bigcup_{\m \geq 2} (\mathcal{A} \times \Al^{\m \times \n}  )$ (see Section \ref{sec: um}).

\begin{exampleon}
To illustrate the problem and notation, let us start with a simple example of model (\ref{modeld}), where $\m = 2$ and the alphabet is binary with $\Al = \{0,1\}$. This means that we consider mixing vectors of the form $\wei = (\wei_1, \wei_2)$ with $  \wei_1 , \wei_2 > 0$, $\wei_1 + \wei_2 = 1$ and two different sources $\Sou_{1\cdot} = (\Sou_{1 1},\ldots,\Sou_{1 \n})$, $\Sou_{2 \cdot} = (\Sou_{21},\ldots,\Sou_{2\n})$ with $\Sou_{\im \iN} \in \{0,1\}$ for $\im = 1,2$ and $\iN = 1,\ldots,\n$. The question we would like to answer is, under which conditions on $\wei$ and $\Sou$ is $(\wei,\Sou)$ uniquely determined via $\mix = \wei \Sou  $.

For a given observation $\mix_\iN$ the underlying source vector $\Sou_{\cdot\iN} = (\Sou_{1\iN}, \Sou_{2\iN})^\top$ equals one of the four different values
\begin{align}\label{exampleEq1}
(0,0)^\top, (1,0)^\top, (0,1)^\top, (1,1)^\top
\end{align}
and hence,
\begin{align}\label{exampleEq2}
\mix_\iN \in \{0, \wei_1, \wei_2, 1\}.
\end{align} 
Clearly, if any two of the four values in the set on the r.h.s. of (\ref{exampleEq2}) coincide, then two different source values in (\ref{exampleEq1}) lead to the same mixture value for $\mix_\iN$ and hence the sources are not identifiable, i.e., they cannot be distinguished. Consequently, a necessary condition for identifiability is that all values in the r.h.s. of (\ref{exampleEq2}) are different, which is equivalent to 
\begin{align}\label{exampleEq3}
\wei_1 \neq \wei_2.
\end{align}
In other words, it is necessary that the alphabet values in $\Al^\m$ are well separated via the mixing weights $a\in \mathcal{A}$. A generalization of this argument to arbitrary alphabets and number of sources is done later in (\ref{emptyi}). Further, we may assume w.l.o.g.\ $\wei_1 < \wei_2$, i.e., we denote that source as $\Sou_{1 \cdot}$ which comes with the smaller weight.
 
(\ref{exampleEq3}) alone, however, is necessary but not sufficient for identifiability. For instance, if $\Sou_{1\iN} = \Sou_{2 \iN}$ for all $\iN = 1,\ldots,\n$ then $\mix_\iN \in \{0,1\}$ and hence, $\wei$ is not identifiable from $\mix$. Thus, a certain variability of the two sources $\Sou_{1\cdot}$ and $\Sou_{2\cdot}$ is necessary to guarantee identifiability of $\wei$.
In this simple example, it is easy to check that a necessary and sufficient variability of $\Sou_{1\cdot}$ and $\Sou_{2\cdot}$ is that $\Sou$ either takes the value $(1,0)$ (i.e., $\mix_\iN = \wei_1$ for some $\iN$) or $(0,1)$ (i.e., $\mix_\iN = \wei_2$ for some $\iN$) as by (\ref{exampleEq3}) and $\wei_1 + \wei_2 = 1$ it follows that $0< \wei_1 < 1/2 < \wei_2 < 1$. In other words, it is necessary that the mixing weight $\wei_1$ (or $\wei_2$ respectively) is seen somewhere in the mixture $\mix$ on its own, without the influence of the other mixing weight. A generalization of these assumptions and this argument to general systems (\ref{model}) is done later in Theorem \ref{theoAlgo} and Theorem \ref{theoAlgoAbw}, respectively.

\end{exampleon}

\section{A Combinatorial Characterisation of Identifiability: General Theory}\label{subsec:nsc}

In model (\ref{modeld}) every observation $\mix_\iN$, for $\iN=1,\ldots,\n$, is given by a linear combination of $\wei \in \mathcal{A}$ (unknown) with one of the finitely many vectors in $\Al^\m$.
So in order to identify $\Sou_{\cdot \iN}$, we have to determine the corresponding vector in $\Al^\m$. 
Note that multiple observed values leave this identification problem invariant, i.e., do not contribute further to identifiability. Hence, w.l.o.g. we assume all observations $\mix_1,\ldots,\mix_\n$ to be pairwise different. Note, that this implies $\n \leq k^\m = \abs{\Al^\m}$.

Of course, when for a given mixing vector $\wei \in \mathcal{A}$ there exist $w^{\prime} \neq w^{\prime \prime} \in \Al^\m$ with $\wei w^{\prime} = \wei w^{\prime \prime}$ and one observes this value, it is not possible to identify the underlying sources $\Sou_{1 \cdot},\ldots,\Sou_{\m \cdot}$ uniquely.
Consequently, a necessary condition for identifiability is that those values are not observed, i.e., $\mix_\iN \notin \{\wei w:\;\exists w^{\prime}\neq w\in\Al^{ \m}\text{ s.t. }\wei w=\wei w^{\prime}\}$ for all $\iN=1,\ldots,\n$. For arbitrary sources this is comprised in the condition of a positive \textit{alphabet separation boundary}, i.e., 
\begin{align}\label{emptyi}
ASB(\wei) \ZuWeis \min_{w^{\prime}\neq w\in\Al^{\m}} \abs{\wei w - \wei  w^{\prime}} > 0.
\end{align}

Let $S_\m^\n$ be the collection of injective maps from $\{1,\ldots,\m\}$ to $\{1,\ldots,\n\}$, i.e., for $\rho \in S_\m^\n$ the vector $(\mix_{\rho(1)},\ldots,\mix_{\rho(\m)})$ corresponds to a selection of elements from $(\mix_1,\ldots,\mix_\n)$.
\begin{theo}\label{theoId}
Assume model (\ref{modeld}) with $\mix = (\mix_1,\ldots,\mix_\n) = \wei \Sou ; $ $(\wei, \Sou) \in \mathcal{A} \times \Al^{\m \times \n}$. Let $E \in \Al^{\m \times \m}$ be an arbitrary but fixed invertible $\m \times \m$ matrix with elements in $\Al$.
Assume that $ASB(\wei) > 0 $ and
\begin{assum}\label{asstheoID}
\newline
there exists $\rho \in S_\m^\n$ such that $(\Sou_{1 \rho(r) }, \ldots, \Sou_{\m \rho(r)})^\top_{1\leq r \leq \m} = E$. 
\end{assum}
Then $(\wei,\Sou)$ is identifiable if and only if
\begin{assum}\label{condit}
\newline
there exists exactly one $\sigma \in S_\m^\n$ such that for $\tilde{\wei} \ZuWeis (\mix_{\sigma(1)},\ldots,\mix_{\sigma(\m)})E^{-1}$ 
\begin{align}\label{condWeiTilde}
\tilde{\wei} \in \mathcal{A} \quad 
\text{ and } \quad
\{\mix_1,\ldots,\mix_\n\} &\in \{\tilde{\wei} w: w\in\Al^\m\},
\end{align}
i.e., $\tilde{\wei}$ is a valid mixing weight and can reproduce all observations.
\end{assum}
\end{theo}
For a proof see Appendix \ref{app:A}.

Theorem \ref{theoId} is fundamental for the following as A\ref{condit} provides a necessary and sufficient condition for identifiability of $(\wei, \Sou)$, given A\ref{asstheoID} holds. In Section \ref{subsec:sc}, it will serve to derive a simple sufficient identifiability condition which is easy to check. 

Theorem \ref{theoId} is formulated for a fixed (but arbitrary) invertible matrix $E$ and the two identifiability conditions A\ref{asstheoID} and A\ref{condit} depend on this matrix $E$ in the following way: For a given $E$ imposing the conditions A\ref{asstheoID} and A\ref{condit} restricts the  set of all possible mixtures $ \mathcal{A} \times \Al^{\m \times \n}$ to a smaller set that depends on $E$ in which all elements are identifiable. Thus, different choices of $E$ lead to different instances of Theorem \ref{theoId}, i.e., to different identifiable submodels. 
In the following, we will discuss the role of $E$ and conditions A\ref{asstheoID} and A\ref{condit} more detailed. 

First, we consider A\ref{asstheoID}.
Assumption A\ref{asstheoID} says that the columns of the fixed matrix $E$ must appear somewhere in the columns of the sources $\Sou$. However, knowledge of where these columns of $E$ occur is not assumed, $\rho$ can be an arbitrary map in $S_\m^\n$. Thus, A\ref{asstheoID} restricts the set of all sources to those where a given set of alphabet combinations, namely the columns of $E$, appears somewhere in the signal. Hence, in practice, it requires pre-knowledge that certain combinations of values in $\Al^\m$ are present \textit{somewhere} in the sources.

Without further restrictions on the matrix $E$, A\ref{asstheoID} simplifies to $\rank(\Sou) = \m$ (in particular implying $\n \geq \m$), which is, indeed, an almost minimal condition. By simple linear algebra, it is easy to check that $\rank(\Sou) < \dim( \spa( \mathcal{A})) = \m-1$ implies that for any $\wei \in \mathcal{A}$ exists an $\tilde{\wei} \neq \wei \in \mathcal{A}$ such that $\wei \Sou = \tilde{\wei} \Sou $, i.e., $(\wei, \Sou)$ is not identifiable. When we allow for arbitrary mixing weights (see Section \ref{sec:abw}), i.e., not necessarily summing up to one, then by the same argument $\rank(\Sou) = \m$ becomes even a necessary condition.
Intuitively, A\ref{asstheoID} ensures that the sources $\Sou_{1\cdot},\ldots,\Sou_{\m \cdot}$ differ sufficiently such that one can identify $\wei$ from their mixture. For instance, if $\Sou_{1 \cdot} = \ldots=\Sou_{\m \cdot}$, it follows that $\mix=\Sou_{1 \cdot}$, irrespective of $\wei$. 
Note that if $k^\m$ different values $\mix_1,\ldots,\mix_\iN$ are observed, i.e., $\n = k^\m$ (recall that w.l.o.g. in this section $\n$ equals the number of pairwise different $\mix_\iN$'s), then it must hold true that $\{\mix_1,\ldots,\mix_\iN\} = \{\wei w: w\in\Al^\m\}$. Thus, A\ref{asstheoID} follows trivially for any invertible $\m \times \m$ matrix $E$ with elements in $\Al$. However, A\ref{asstheoID} is a much weaker assumption than $\n = k^\m$. 

Second, we comment on assumption A\ref{condit}. Given A\ref{asstheoID}, assumption A\ref{condit} reveals $\wei$ as identifiable as soon as we can assign a collection of observations to rows in $E$ in a unique way. If for some $\sigma \neq \rho \in S_\m^\n$ $\tilde{\wei}$ in A\ref{condit} fulfills the conditions in (\ref{condWeiTilde}), $\tilde{\wei}$ is a different mixing weight which can produce the same mixture $\mix$ with some sources fulfilling A\ref{asstheoID} and hence, $(\wei, \Sou)$ is not identifiable. However, if such a $\sigma \neq \rho \in S_\m^\n$ does not exists, $\wei$ is uniquely determined from $\mix$ and identifiability of $\Sou$ follows from $ASB(\wei) > 0$. 


In Section \ref{subsec:sc} we will show that for some specific choices of $E$ A\ref{condit} always holds, i.e., A\ref{asstheoID} already implies identifiabiliy. The following example shows that this is not true in general, i.e., not for any choice of $E$.

\begin{exampleon}\label{gsi}
With the notation of (\ref{modeld}) and Theorem \ref{theoId}, let $\n=\m=3$, $\Al=\{0,1,\frac{21}{6+\sqrt{15}},6\}$, $\wei = \left(\frac{6 - \sqrt{15}}{30}, \frac{6 + \sqrt{15}}{30}, 0.6 \right)$, and 
\begin{align*}
\Sou = E = \begin{pmatrix} 6 & 0 & 0 \\ 0 & \frac{21}{6+\sqrt{15}} & 0 \\ 0 & 0 & 1\end{pmatrix},
\end{align*}
i.e., $\rho$ in A\ref{asstheoID} is the identity map and
\begin{align*}
\mix = \wei \Sou  = \left(\frac{6-\sqrt{15}}{5} , 0.7 , 0.6 \right).
\end{align*}

For $\sigma: (1,2,3) \mapsto (3, 1, 2)$ we find that 
\begin{align*}
\left( \mix_{\sigma (1)} , \mix_{\sigma(2)} , \mix_{\sigma(3)} \right) E^{-1}  = \left( 0.1 , 0.2 , 0.7 \right) \WeisZu \tilde{\wei},
\end{align*}
which is a valid mixing weight. Hence, $(\wei, \Sou)$ is not identifiable.
\end{exampleon}

As mentioned before, in Section \ref{subsec:sc} we will show that some specific choices of $E$ already lead to uniqueness of the selection $\sigma$ in A\ref{condit}, and thus ensure identifiability. The following remark illustrates how specific choices of rows in $E$ fix some subdomain of $\sigma$ in A\ref{condit}. 

\begin{remark} \label{remark1}
If $\n = k^\m$ with $\mix_1<\ldots<\mix_{k^\m}$, then $\mix_1$, the smallest observed value, corresponds to the situation when all sources $\Sou_{1 \iN},\ldots,\Sou_{\m \iN}$ take the smallest value of the alphabet (denoted with $\al_1$), i.e., if $E_{1 \cdot} = (\al_1,\ldots,\al_1)$, then for all $\sigma$ satisfying A\ref{condit} $\sigma(1) = 1$.
The second smallest observed value, $\mix_2$, corresponds to the situation when all sources $\Sou_{2 \iN},\ldots,\Sou_{\m \iN}$ take the smallest value $\al_1$, but the source $\Sou_{1 \iN}$ with the smallest weight $\wei_1$ takes the second smallest value $\al_2$, i.e., if $E_{2 \cdot} = (\al_2,\al_1,\ldots,\al_1)$, then for all $\sigma$ satisfying A\ref{condit} $\sigma(2) = 2$. Analogous holds for the largest observed value and the second largest observed value.
\end{remark}

\section{A Simple Sufficient Identifiability Criterion}\label{subsec:sc}

We have seen in Theorem \ref{theoId} that the problem of identifying $\wei$ and $\Sou$ from the observations $\mix_1,\ldots,\mix_\n$ reduces to find the corresponding observations $\mix_\iN$ for $\m$ linear independent rows of an invertible $\m \times \m$ matrix $E$ with elements in $\Al$.
Remark \ref{remark1} points out that some observations $\mix_\iN$ can always be uniquely assigned to source vectors $(\Sou_{1 \iN},\ldots,\Sou_{\m \iN})^\top\in\Al^\m$ and thus, limit the possible maps $\sigma$ in A\ref{condit}. The next theorem shows that there is even more structure in the observations $\mix_1,\ldots,\mix_\n$ and that certain variations in the sources, i.e., certain choices of $E$ in A\ref{asstheoID}, already ensure identifiability.
Moreover, the proof of the following theorem gives an explicit construction of the unique $(\wei, \Sou)$ from $\mix$. 

\begin{theo}\label{theoAlgo}
Assume model (\ref{modeld}) with $\mix = (\mix_1,\ldots,\mix_\n) = \wei \Sou ; $ $(\wei, \Sou) \in \mathcal{A} \times \Al^{\m \times \n}$.
Furthermore, assume that $ASB(\wei) > 0$ and
\begin{assum}\label{assusuff}
there exists $\rho \in S_\m^\n$ such that 
\begin{align*}
\left(\Sou_{1 \rho(r)}, \ldots, \Sou_{\m \rho(r)}\right)^{\top} = \sAl^r, \qquad r = 1, \ldots, \m,
\end{align*}
with $(\sAl^r_\im)_{1\leq \im \leq \m} \ZuWeis \left( \al_1\mathds{1}_{\im \neq r} + \al_2 \mathds{1}_{\im = r}\right)_{1\leq \im \leq \m}$. 
\end{assum}

Then $(\wei,\Sou)$ is identifiable.
\end{theo}

Before we give a proof (which is based on an explicit algorithm to compute $\wei$ from $\mix_1,\ldots,\mix_\n$) we will discuss relationships and differences between the previous results and their assumptions A\ref{asstheoID} - A\ref{assusuff}.

Assumption A\ref{assusuff} of Theorem \ref{theoAlgo} has a simple interpretation. It means that each of the mixing weights $\wei_\im$ appears somewhere in the mixture on its own (without the influence of any other mixing weight $\wei_\imp \neq \wei_\im$) via the mixture value $\mix_{\rho(\im)} = \wei \sAl^\im = (\al_2-\al_1) \wei_\im + \al_1$. For instance, if the alphabet is of the form $\Al = \{0,1,\al_3,\ldots, \al_k\}$ A\ref{assusuff} simplifies to the condition that the mixing weights appear somewhere in the mixture, i.e., $\wei_\im \in \{\mix_1,\ldots,\mix_\iN\}$ for all $\im = 1,\ldots,\m$. Intuitively, A\ref{assusuff} means that for each $\im = 1,\ldots,\m$ there exists one mixture observation $\mix_\iN$ such that only $\Sou_{\im \iN}$ is active (taking the value $\al_2$) and all other sources $\Sou_{\imp \iN}$ with $\imp \neq \im$ are silent (taking the value $\al_2$). 
It is easy to check that the choice of alphabet values $\al_1$ and $\al_2$ (the smallest and second smallest alphabet value) in Theorem \ref{theoAlgo} can be replaced by $\al_k$ and $\al_{k-1}$ (the largest and second largest alphabet value).

Obviously, assumption A\ref{assusuff} arises from assumption A\ref{asstheoID} for a specific choice of $E$ in Theorem \ref{theoId}, namely with
\begin{align}\label{faMatrix}
E = \begin{pmatrix}
\sAl^1 \\
\vdots \\
\sAl^m
\end{pmatrix} =
\begin{pmatrix}
\al_2 & \al_1 & \ldots & \al_1 & \al_1 \\
\al_1 & \al_2 & \ldots & \al_1 & \al_1\\
\al_1 & \al_1 & \ddots & \al_1 & \al_1 \\
\al_1 & \al_1 & \ldots & \al_2 & \al_1 \\
\al_1 & \al_1 & \ldots & \al_1 & \al_2
\end{pmatrix}.
\end{align}
Consequently, if the matrix in (\ref{faMatrix}) is invertible, A\ref{assusuff} implies A\ref{asstheoID}. The following Lemma shows that this holds under mild conditions on $\Al$ and $\m$.

\begin{lemon}\label{AA}
For model (\ref{modeld}) let A\ref{assusuff} be as in Theorem \ref{theoAlgo} and A\ref{asstheoID} as in Theorem \ref{theoId}.\\
If $\al_2\neq \al_1(1-\m)$, then A\ref{assusuff} implies A\ref{asstheoID}. 
\end{lemon}

For a proof see Appendix \ref{app:B}.

Figure \ref{Arelation} summarizes all relations between A\ref{asstheoID} - A\ref{assusuff} in a diagram.

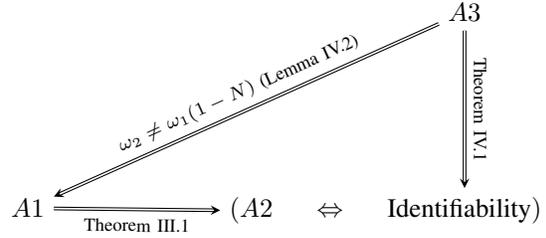
\begin{figure}[h]
\begin{tikzpicture}
  \matrix (m) [matrix of math nodes,row sep=1em,column sep=1em,minimum width=1em]
  {
     \quad &\quad &\quad & \quad &\quad & A3 \\
     \quad &\quad & \quad & \quad & \quad &  \quad \\
      \quad &\quad & \quad & \quad & \quad &  \quad \\
       \quad &\quad & \quad & \quad & \quad &  \quad \\
     A1 & \quad &\quad & (A2 &\Leftrightarrow& \text{Identifiability}) \\};
  \path[-stealth]
    (m-1-6) edge [double] node[sloped,midway,above] {\scriptsize{$\al_2\neq \al_1(1-\m)$ (Lemma \ref{AA})}}  (m-5-1)
    (m-1-6) edge [double] node[sloped,midway,above] {\scriptsize{Theorem \ref{theoAlgo}}}  (m-5-6)
    (m-5-1) edge [double] node[sloped,midway,below] {\scriptsize{Theorem \ref{theoId}}}  (m-5-4);
\end{tikzpicture}
\caption{Relation between  A\ref{asstheoID}, A\ref{condit}, and A\ref{assusuff} }
\label{Arelation}
\end{figure}

Now we turn to the proof of Theorem \ref{theoAlgo}, which is proven by explicit recovery of $(\wei, \Sou)$.
This generalizes an algorithm of Diamantaras and Chassiot \cite{diamantras2} for the binary alphabet $\Al = \{-1,1\}$ to a general finite alphabet.

\begin{IEEEproof}[Proof of Theorem \ref{theoAlgo}]
Assumption A\ref{assusuff} implies that 
\begin{align}\label{defcr}
\mix_{\rho(r)} = \wei \sAl^r, \quad r=1,\ldots,\m,
\end{align}
and hence,
\begin{align}\label{remarkOmega}
\wei_r = \frac{\mix_{\rho(r)}-\al_1}{\al_2 - \al_1}.
\end{align}
Thus, it suffices to determine the map $\rho$ and the values $\mix_{\rho(1)},\ldots,\mix_{\rho(\m)}$, respectively, in order to determine $\wei$. When $\wei$ is determined, identifiability of $\Sou$ follows from $ASB(\wei) > 0$.

Recall Remark \ref{remark1} and note that 
\begin{align}\label{remarkC1}
\mix_{\rho(1)} = \min(\{\mix_1,\ldots,\mix_\n\} \backslash \{\al_1\}),
\end{align}
which determines $\wei_1$ as in (\ref{remarkOmega}).
The following two lemmas show that successively all the other $\rho(r)$ (and hence $\wei_r$) for $r = 2,\ldots,\m$ can be determined as well, which finishes the proof. 

Let $B=B(\m,\Al)$ be the $\m \times k^\m$-matrix, where the $j$th column of $B$ is the number $j-1$ written in the positional notation based on the number $k$, identifying $0$ with $\al_1$, $1$ with $\al_2$, and so on, i.e.,

\begin{align}\label{Bmatrix}
B_{ij}= \sum_{r=1}^k \al_r \mathds{1}_{\lfloor \frac{(j-1) \Mod{ k^{i}}}{k^{i-1}}\rfloor =r-1},
\end{align}
and $d \ZuWeis \wei B$ the $k^\m$ dimensional vector of all possible values that $\mix$ can take.

\begin{lemon}\label{lemmaD}
From $\wei_{1},\ldots,\wei_{r}$ one can determine $d_1,\ldots,d_{k^{r}}$ uniquely.
\end{lemon}

\begin{lemon}\label{lemmaCR}
It holds that
$\mix_{\rho(r)} = \min(\{\mix_1,\ldots,\mix_\n\} \backslash \{d_1,\ldots, d_{k^{r-1}}\})$.
\end{lemon}

For proofs see Appendix \ref{app:Bb} and \ref{app:C}, respectively.

\end{IEEEproof}

\begin{figure}[h!]
\begin{algorithmic}
\Input $\mix_1,\ldots,\mix_\n$
\State $r=1$
\State $\fG \gets \{\mix_1,\ldots,\mix_\n\}\setminus \{\al_1\}$
\State $c_1\gets \min \fG$
\State $\wei_1 \gets \frac{c_1 - \al_1}{\al_2-\al_1}$
\State Determine $d_1,\ldots,d_{k}$ with  Lemma \ref{lemmaD} (using $\wei_{1}$).
\State $\fG \gets \fG \setminus \{d_1,\ldots,d_{k}\}$
\While{ $\fG \neq \emptyset$}
	\State $r=r+1$
	\State $c_{r}\gets \min \fG$
	\State $\wei_r \gets \frac{c_r - \al_1}{\al_2-\al_1}$
	\State Determine $d_1,\ldots,d_{k^{r}}$ with  Lemma \ref{lemmaD} (using $\wei_{1},\ldots,\wei_{r}$).
	\State $\fG \gets \fG \setminus \{d_1,\ldots,d_{k^{r}}\}$
\EndWhile
\State $\m\gets r$\\
\Return $\m$ and $\wei_1,\ldots, \wei_\m$ 
\end{algorithmic}
\caption{
Algorithm for weight identification in (\ref{modeld}), with $\al_1<\ldots<\al_k$, under assumption A\ref{assusuff} of Theorem \ref{theoAlgo}.} \label{fig:algodetmu}
\end{figure}

The proof of Theorem \ref{theoAlgo} gives an explicit recovery construction for $(\wei, \Sou)$ which is summarized in Figure \ref{fig:algodetmu}.
For noisy data one may use algorithm \ref{fig:algodetmu} to proceed similar as Diamantaras and Chassiot \cite{diamantras2} who suggest a clustering approach for estimating $\wei$ and $\Sou$ from noisy observations of $\mix$. However, as the purpose of this paper is not to propose a practical method for recovery from noisy data, but rather to analyze the scope of scenarios under which this is possible in principle, we are not going to follow this approach here further.

\section{Identifiability for Mixtures of Stochastic Processes}\label{sec:sp}

In this subsection we will shortly discuss how likely it is for the identifiability condition of Theorem \ref{theoAlgo} to be satisfied when $(\Sou_{1 \iN},\ldots,\Sou_{\m \iN})^\top_\iN$ is a stochastic process. Therefore, let $\sAl^r$ be as in Theorem \ref{theoAlgo}, and define the hitting times
\begin{align}\label{Tr}
\fT^r\ZuWeis \min\{\iN\in \N: (\Sou_{1 \iN},\ldots,\Sou_{\m \iN})=\sAl^r\},
\end{align}
for $r=1,\ldots,\m$, and the stopping time
\begin{align*}
\fT \ZuWeis \max_{r=1,\ldots,\m}\fT^r.
\end{align*}
Then it follows from Theorem \ref{theoAlgo} that
\begin{align}\label{poi}
\begin{aligned}
&\Pp\left((\wei,\Sou)\text{ is identifiable}\right)\\
\geq &\Pp\left(\exists \rho \in S_\m^\n : (\Sou_{1  \rho(r)}, \ldots,\Sou_{\m \rho(r)})^{\top}  = \sAl^r\right)\\
= &\Pp\left(\fT \leq \n\right)
\geq 1-\sum_{r=1}^\m \Pp\left(\fT^r > \n \right).
\end{aligned}
\end{align}
Note that this bound only depends on the distributions of the hitting times $\fT^r$, which are often explicitly known or good estimates exist.
A prominent class of examples for modeling the distribution of the source signals are Markov processes including iid sequences (see e.g., \cite{yau,diamantaras4,rostami}).

\begin{theo}\label{theoexp}
Assume that the source signals $(\Sou_{1 \iN},\ldots,\Sou_{\m \iN})^\top_\iN$ in (\ref{modeld}) constitute an irreducible Markov process on the finite state space $\Al^\m$, with transition matrix $P = (p_{ij})_{1\leq i,j \leq k^\m}$, where we identify the first $\m$ states of $\Al^\m$ with $\sAl^1, \ldots, \sAl^\m$ from Theorem \ref{theoAlgo}. Let $\Nn \in \N$ be such that $P^\Nn > 0$ and
\begin{align*}
Q_r &\ZuWeis (p_{ij})_{1\leq i,j \neq r \leq k^\m}, \qquad r = 1, \ldots, \m, \\
c &\ZuWeis \max_{1\leq r \leq \m}\norm{Q_r^\Nn \mathbf{1}}_{\infty}.
\end{align*}
Then $c < 1$ and, if $ASB(\wei) > 0$, 
\begin{align}\label{pIexpB}
1-\Pp((\wei,\Sou)\text{ is identifiable}) \leq \m c^{\lfloor \frac{\n}{\Nn}\rfloor} \leq \m c^{\frac{\n - \Nn}{\Nn}}.
\end{align}
\end{theo}

For a proof see Appendix \ref{app:E}.

\begin{example}{Bernoulli Model}\label{exB}
Let us consider (\ref{modeld}) for the simple case where we have two sources $\Sou_{1 \cdot}$ and $\Sou_{2 \cdot}$ that can take two different values, i.e., $\Al = \{\al_1,\al_2\}$. For instance, the source signals could come from a binary antipodal alphabet ($\Al=\{-1,1\}$) as they appear in many digital modulated schemes.

If we assume that $\Sou_{1 \iN}$ and $\Sou_{2 \iN}$ are independent and identically distributed (i.i.d.) for all $\iN = 1,\ldots, \n$ with $\Pp(\Sou_{\im \iN} = \al_1) = p \in (0,1)$ and $\Pp(\Sou_{\im \iN} = \al_2) = 1- p \WeisZu q$ for $\im = 1,2$ and $\iN = 1,\ldots,\n$, then $(\Sou_{1 \iN}, \Sou_{2 \iN})^\top_\iN$ constitutes an irreducible Markov process on the state space $\{(\al_2, \al_1)^\top, (\al_1, \al_2)^\top, (\al_1, \al_1)^\top, (\al_2, \al_2)^\top \}$ with transition matrix
\begin{align*}
P = \begin{pmatrix}
pq & pq & p^2 & q^2 \\
pq & pq & p^2 & q^2 \\
pq & pq & p^2 & q^2 \\
pq & pq & p^2 & q^2 
\end{pmatrix} > 0.
\end{align*}
Hence, $\Nn = 1$, 
\begin{align*}
Q_1 = Q_2 = \begin{pmatrix}
pq & p^2 & q^2 \\
pq & p^2 & q^2 \\
pq & p^2 & q^2
\end{pmatrix},
\end{align*}
and $c = qp + p^2 + q^2 = 1 - pq$. Thus, Theorem \ref{theoexp} yields
\begin{align*}
1 - \Pp((\wei,\Sou)\text{ is identifiable}) \leq  2(1 - pq)^\n.
\end{align*}
In this simple setting we can even calculate the probability of identifiability exactly. Note that $(\wei,\Sou)$ is identifiable if and only if $c_1 = \wei_1 \al_2 + \wei_2 \al_1$ or $c_2 = \wei_1 \al_1 + \wei_2 \al_2$ is observed as $\al_1 < c_1 < \al_1 + (\al_2 - \al_1)/2 < c_2  < \al_2$ and 
\begin{align*}
(\wei_1,\wei_2)&= \left(\frac{c_1 - \al_1}{\al_2 - \al_1},\frac{\al_2-c_1}{\al_2-\al_1}\right)\\
&= \left(\frac{\al_2 - c_2}{\al_2 - \al_1},\frac{c_2-\al_1}{\al_2-\al_1}\right).
\end{align*}
Therefore,
\begin{align*}
\begin{aligned}
1 - \Pp((\wei,\Sou)\text{ is identifiable})
= \Pp(\Sou_{1 1} = \Sou_{21})^\n =  (1 - 2pq)^\n.
\end{aligned}
\end{align*}
\end{example}
Example \ref{exB} shows that the bound in Theorem \ref{theoexp} does not need to be sharp in general but captures the exponential decay (in $\n$) well. This is mainly because in Theorem \ref{theoexp} the probability of $(\wei, \Sou)$ being identifiable is bounded using the sufficient (and not necessary) identifiability condition A\ref{assusuff} from Theorem \ref{theoAlgo}. In Section \ref{sec:sim} the gap between this bound and the true probability $\Pp((\wei,\Sou)\text{ is identifiable})$ in (\ref{pIexpB}) is further explored in a simulation study.

\section{Multiple Linear Mixtures}\label{sec:mlm}
After analyzing the most difficult scenario of a single linear mixture with $\M = 1$ in (\ref{model}), generalizations to arbitrary number of mixtures $\M$ now follow easily.
To this end, for a vector $v$ let $\norm{v}_1\ZuWeis \sum_{i=1}^n \abs{v_i}$ denote the $l_1$-norm and define the set of $\M$-mixtures as
\begin{align*}
\mathcal{A}_\M \ZuWeis \{\Wei \in \R_+^{\M \times \m}:\norm{\Wei_{\cdot 1}}_1 < ... < \norm{\Wei_{\cdot 1}}_1,\\
  \norm{\Wei_{1 \cdot}}_1 =...= \norm{\Wei_{\M \cdot}}_1 = 1\}.
\end{align*}
Again, note that $\norm{\Wei_{\cdot \im}}_1 \neq \norm{\Wei_{\cdot \imp}}_1 $ for $\im \neq \imp$ is necessary to ensure identifiability of different sources.
Then for $\Sou \in \Al^{\m \times \n} $ and 
 $\Wei \in \mathcal{A}_\M$ the observed values are given by $\Mix = \Wei \Sou $, i.e.,
\begin{align*}
\Mix_{ \iM \iN} = \sum_{\im = 1}^\m \Wei_{\iM \im} \Sou_{\im \iN}, \qquad \iN = 1,\ldots,\n, \; \iM = 1,\ldots,\M.
\end{align*}
Identifiability means to decompose the matrix $\Mix \in \R^{\M \times \n}$ uniquely into matrices $\Wei  \in \mathcal{A}_P$ and $\Sou \in \Al^{\m \times \n}$
for given finite alphabet $\Al = \{ \al_1,\ldots,\al_k \}$ with $\al_1 < \ldots < \al_k$ and given $\n, \m, \M \in \N$.
Analog to before we define the alphabet separation boundary of a mixture matrix $\Wei \in \mathcal{A}_\M$ as
\begin{align*}
ASB(\Wei) \ZuWeis \min_{w^{\prime}\neq w\in\Al^{\m}} \norm{\Wei w - \Wei  w^{\prime}}_1.
\end{align*}
Clearly, $ASB(\Wei) > 0$ is a necessary condition on $\Wei$ for $(\Sou, \Wei)$ to be identifiable.
Theorem \ref{theoId}, Theorem \ref{theoAlgo}, and Theorem \ref{theoexp} assume that $\M=1$.
It is straight forward to check that Theorem \ref{theoId} holds unchanged when $\M > 1$ with $\mathcal{A}$ replaced by $\mathcal{A}_P$.
The same is true for Theorem \ref{theoAlgo}, where in the proof the minimum of a set of observations $\Mix_{\cdot \iN}$ must be replaces by the minimum defined in terms of the ordering of $\sum_{\iM = 1}^\M \Mix_{\iM \iN }$.
Thus, clearly Theorem \ref{theoexp} also holds unchanged when $\M > 1$.


\section{Arbitrary mixing weights}\label{sec:abw}

So far, we assumed the mixing weights to be positive and to sum up to one. However, in some applications this assumption is not satisfied (e.g., in digital communications \cite{proakis}) and in the following we discuss such generalizations. 
Let  $\tilde{\mathcal{A}}  \subseteq \mathcal{A}_0 \ZuWeis \{ \wei \in \R^\m: \wei_1 < \ldots < \wei_\m \} $ be an arbitrary subset of mixing weights $\wei_\im \in \R$. Note that w.l.o.g. $\wei_1 < \ldots < \wei_\m$ in order to assign the mixing weight to a source.

It is easy to check that Theorem \ref{theoId} holds unchanged with $\mathcal{A}$ replaced by $\tilde{\mathcal{A}}$. Note, however, that if $\tilde{\mathcal{A}} \supsetneq \mathcal{A}$ condition A\ref{condit} becomes more restrictive, i.e., a mixture $(\wei, \Sou)$ which is identifiable with respect to $\mathcal{A}$ might not be identifiable with respect to $\tilde{\mathcal{A}}$.

Analogously, Theorem \ref{theoAlgo} can be generalized for  $\tilde{\mathcal{A}} \supsetneq \mathcal{A}$, where now the corresponding identifiability assumption A\ref{assusuff} becomes more restrictive.

The following theorem considers the most general case of arbitrary mixing weights in $\mathcal{A}_0$.
\begin{theo}\label{theoAlgoAbw}
Assume model (\ref{modeld}) $\mix = \wei \Sou  $ with  $(\wei, \Sou) \in \mathcal{A}_0 \times \Al^{\m \times \n}$.
Furthermore, assume that $ASB(\wei) > 0$ and there exists $\rho, \mu \in S_{\m + 1}^\n$ such that 
\begin{align}
\begin{aligned}\label{assusuffabw}
\left(\Sou_{1 \rho(r)}, \ldots, \Sou_{\m \rho(r)}\right)^{\top} = \sAl^{r-1}, \; r = 1, \ldots, \m + 1,\\
\left(\Sou_{1 \mu(r)}, \ldots, \Sou_{\m \mu(r)}\right)^{\top} = \sAlv^{r-1}, \; r = 1, \ldots, \m + 1,
\end{aligned}
\end{align}
with $\sAl^{r}, \sAlv^{r} \in \Al^\m$ for $r = 0, \ldots, \m$ defined as 
\begin{align*}
(\sAl^{r})_i &\ZuWeis \al_2\mathds{1}_{i = r}  + \al_k \mathds{1}_{\wei_i * \wei_{r} < 0} + \al_1 \mathds{1}_{\substack{\wei_i * \wei_{r} > 0 \\ i \neq r }}, \\
(\sAlv^{r})_i &\ZuWeis \al_{k-1}\mathds{1}_{i = r} + \al_1\mathds{1}_{\wei_i * \wei_{r} < 0} + \al_k \mathds{1}_{\substack{\wei_i * \wei_{r} > 0 \\ i \neq r }},
\end{align*} 
for $i = 1,\ldots, \m$ and $\wei_0 \ZuWeis 1$.
Then $(\wei,\Sou)$ is identifiable.
\end{theo}
The proof of Theorem \ref{theoAlgoAbw} is given in Appendix \ref{sec: apptheAlgoAbw}.

Recall that for positive mixing weights the identifiability condition A\ref{assusuff} in Theorem \ref{theoAlgo} had a very simple interpretation, namely that each of the single mixing weights $\wei_\im$ appears somewhere in the mixture $\mix$ on its own, without the influence of any of the other mixing weights $\wei_\imp$ for $\imp \neq \im$. The interpretation of (\ref{assusuffabw}) is somewhat more difficult, but similar.
In the case of probability mixing weights $\wei \in \mathcal{A}$ as in Theorem \ref{theoAlgo} both, the sum and the absolute sum of the mixing weights were fixed via $\sum_{\im=1}^\m \wei_\im = \sum_{\im=1}^\m \abs{\wei_\im} = 1$ and this determined the scaling in which the mixing weights appear in the mixture $\mix$. Now for general mixing weights $\wei \in \mathcal{A}_0$ as in Theorem \ref{theoAlgoAbw} both, the sum and the absolute sum (or equivalently the sum of the negative mixing weights and the sum of the positive mixing weights) are unknown and thus, additional conditions to determined these unknown scaling parameters are needed. These correspond to $\sAl^0$ and $\sAlv^0$. They ensure that the smallest possible mixture value (which corresponds to $\sAl^0$) and the largest possible mixture value (which corresponds to $\sAlv^0$) are observed and thus determine the scaling parameters.
Now analog to $\sAl^\im$ in A\ref{assusuff} of Theorem \ref{theoAlgo}, $\sAl^\im$ and $\sAlv^\im$ in (\ref{assusuffabw}) of Theorem \ref{theoAlgoAbw} ensure that $\wei_\im$ appears somewhere in the mixture $\mix$ on its own and can thus be determined. However, as the sign of $\wei_\im$ is now unknown, too, we get the additional unambiguity that a mixture value can be increased either by increasing a source which corresponds to a positive mixing weight or by decreasing a source which corresponds to a negative mixing weight.

From Theorem \ref{theoAlgoAbw} it follows directly that Theorem \ref{theoexp} holds with $\m$ replaced by $2\m+2$, when we allow for arbitrary mixing weights in  $\mathcal{A}_0$.

\section{Unknown number of source components}\label{sec: um}
So far, we assumed that the number of sources $\m$ is fixed and known. Now we consider the case where $\m$ is unknown, i.e. $\mix = \wei \Sou $ with
\begin{align}\label{ddmUm}
(\wei,\Sou) \in \bigcup_{\m\geq 2} (\mathcal{A} \times \Al^{\m \times \n}).
\end{align}
While it is not clear how to generalize Theorem \ref{theoId} for $(\wei,\Sou)$ as in (\ref{ddmUm}), condition A\ref{assusuff} in Theorem \ref{theoAlgo} is still sufficient for identifiability when $\m$ is unknown.

To see this, note that the proof of Theorem \ref{theoAlgo} (in particular Lemma \ref{lemmaD}) does not require the number of sources $\m$ to be known, where $\m$ is determined via
\begin{align*}
\m = \min(r \in \N_{\geq 2} \text{ s.t. } \{\mix_1,\ldots,\mix_\n\} \subset \{d_1,\ldots,d_{k^r}\}),
\end{align*}
and thus, we obtain the following theorem.
\begin{theo}\label{cor: idm}
Assume model (\ref{modeld}) with $\mix = (\mix_1,\ldots,\mix_\n) = \wei \Sou ; $ $(\wei, \Sou) \in \bigcup_{\m \geq 2} (\mathcal{A} \times \Al^{\m \times \n})$.
Furthermore, assume that $ASB(\wei) > 0$ and A\ref{assusuff} holds.
Then $(\wei,\Sou)$ (and thus $\m$) is identifiable.
\end{theo}
Analogously, Theorem \ref{theoexp} and Theorem \ref{theoAlgoAbw} do not require $\m$ to be known.

\section{Simulations}\label{sec:sim}
Finally, we explore in a simulation study how far assumption A\ref{assusuff} from Theorem \ref{theoAlgo} is from being necessary when the sources come from an irreducible Markov process; which corresponds to exploring the tightness of the bound in Theorem \ref{theoexp}.
To this end, Theorem \ref{theoId} is fundamental as it enables us to explicitly examine identifiability of $(\wei, \Sou)$.
We consider an example with a binary alphabet $\Al = \{-2, 1\}$ and a mixture of $\m = 3$ sources.
The matrix $E$ in Theorem \ref{theoId} was chosen randomly over the set of invertible $\m \times \m$ matrices with elements in $\Al$. Simulation runs were always $1,000$.

First, we assume the mixing weights $\wei = (0.2, 0.35, 0.45)$. Note that $\wei  \al \neq \wei \al^{\prime}$ for all $\al \neq \al^{\prime} \in \{-2, 1\}^3$, i.e., $ASB(\wei) > 0$.
\subsection{Bernoulli Model}\label{subsec:simB}
\begin{figure}[h!]
\centering
  \includegraphics[width=\linewidth]{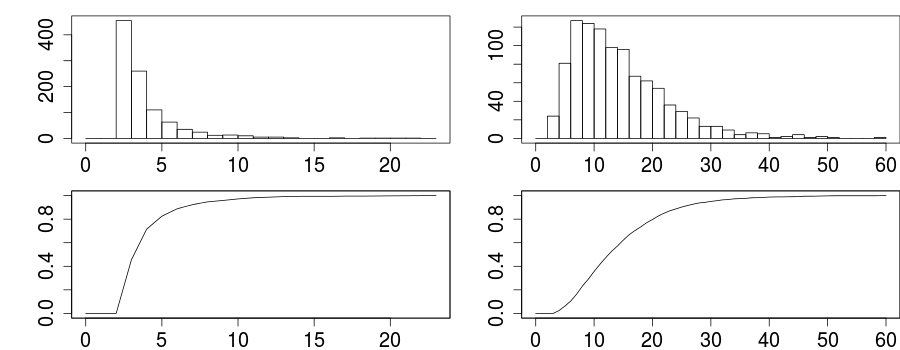}
  \caption{Top row: histograms of number of observations until $(\wei,\Sou)$ is identifiable (left) and until the sufficient identifiability condition from Theorem \ref{theoAlgo} is fulfilled (right),
bottom row: corresponding empirical cumulative distribution function,  
   with $\m = 3$, $\Al = \{-2, 1\}$, $\wei = (0.2, 0.35, 0.45)$ and $\Sou_{\im \iN}$ i.i.d. on $\Al$.}
  \label{fig: id_hist}
\end{figure}

Assuming the sources to be i.i.d. for all $\im = 1,2,3$ and $\iN = 1,\ldots,\n$, with $\Pp(\Sou_{\im \iN} = -2) = \Pp(\Sou_{\im \iN} = 1) = 0.5$ we find that $(\wei,\Sou)$ is already identifiable on average for $\iN \geq 4.39 $ observations. For the sufficient identifiability condition from Theorem \ref{theoAlgo} to hold we find an average value of $\iN \geq 14.79$ observations. Figure \ref{fig: id_hist} shows the corresponding histograms and cumulative distribution functions. The results indicate that the number of observations needed for the sufficient identifiability condition A\ref{assusuff} from Theorem \ref{theoAlgo} to be satisfied is not considerably higher then the actual number of observations until $(\wei,\Sou)$ is identifiable and thus the bound in Theorem \ref{theoexp} is quite sharp in this example.

\subsection{Markov Model}

\begin{figure}[h!]
\centering
  \includegraphics[width=\linewidth]{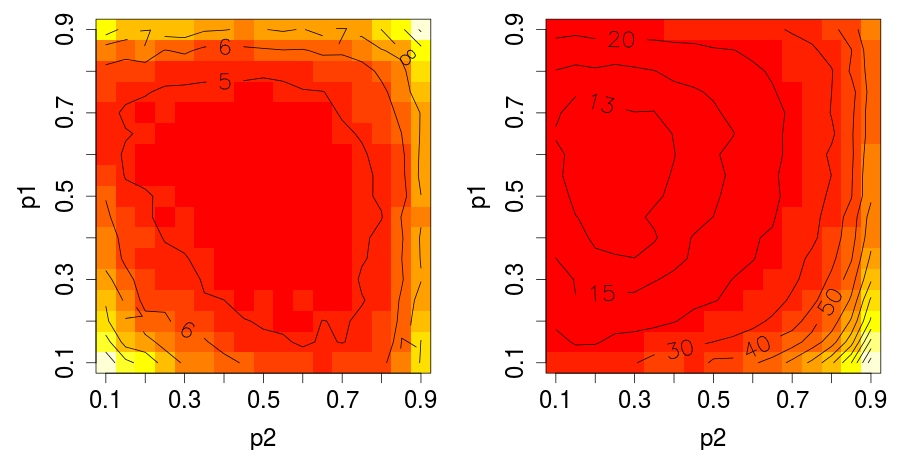}
  \caption{Heat-maps of average number of observations until $(\wei,\Sou)$ is identifiable (left) and until the sufficient identifiability condition from Theorem \ref{theoAlgo} is fulfilled (right) in dependence of $p_1$ and $p_2$ from (\ref{transM}).}
  \label{fig: id_markov}
\end{figure}

We consider a more general Markov model for generating the sources, i.e., we assume the sources to be independent Markov processes on the state space $\Al = \{-2, 1\}$ with transition matrix 
\begin{align}\label{transM}
P = \begin{pmatrix} p_1 & 1-p_1 \\ 1-p_2 & p_2 \end{pmatrix}.
\end{align}
In Figure \ref{fig: id_markov} we display the average numbers of observations until $(\wei, \Sou)$ is identifiable and until the sufficient identifiability condition A\ref{assusuff} from Theorem \ref{theoAlgo} is fulfilled, respectively, for each $(p_1, p_2) \in \{0.1, 0.15, 0.2, \ldots, 0.8, 0.85, 0.9\}^2$. Note that $p_1 = 1 - p_2$ corresponds to i.i.d. observations, with $1 - \Pp(\Sou_{\im \iN} = 1) = \Pp(\Sou_{\im \iN} = -2) = p_1$.

From Figure \ref{fig: id_markov} we draw that identifiability is achieved faster when $p_1$ and $p_2$ are close to $0.5$, which corresponds to the i.i.d. Bernoulli model from Section \ref{subsec:simB}. This is explained by condition A\ref{condit} in Theorem \ref{theoId}, where a richer variation in the sources, i.e., many different observations $\mix_1, \ldots, \mix_\n$, reduces the set of possible valid mixing weights and thus favor identifiability. The sufficient identifiability condition A\ref{assusuff} from Theorem \ref{theoAlgo}, however, requires repeated occurrence of the smallest alphabet values $\al_1$. Consequently, small $p_1$ and large $p_2$ discriminate against those variations.

\subsection{Multiple Linear Mixtures}

Now, we consider multiple linear mixtures, i.e., $\M > 1$. Therefore, for each run, we draw $\M$ mixing weights, each of length $\m = 3$, independently from the uniform distribution on $\mathcal{A}$ (implying $ASB(\wei) > 0$).
For the sources, we consider a Bernoulli model as in Section \ref{subsec:simB}.

\begin{figure}[h!]
\centering
  \includegraphics[width=\linewidth]{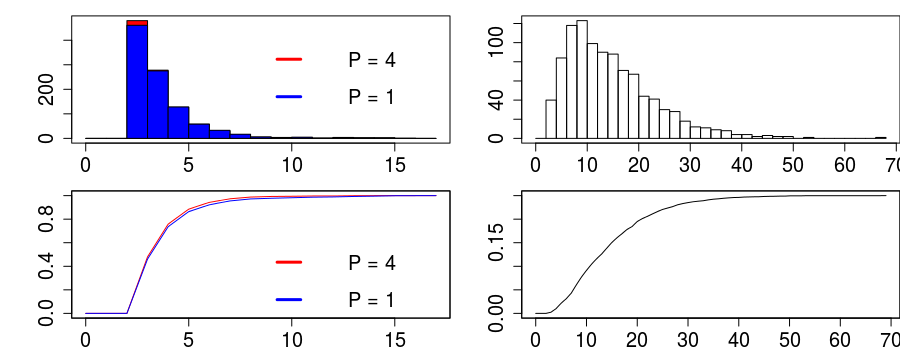}
  \caption{Top row: histograms of number of observations required until $(\wei,\Sou)$ is identifiable (left) and until the sufficient identifiability condition A\ref{assusuff} from Theorem \ref{theoAlgo} is fulfilled (right),
bottom row: corresponding empirical cumulative distribution function,  
with $\m = 3$, $\Al = \{-2, 1\}$, $\wei$ uniformly distributed on $\mathcal{A}$ and $\Sou_{\im \iN}$ i.i.d. on $\Al$.}
\label{fig: id_hist_mlm}
\end{figure}

We find that $(\wei,\Sou)$ is identifiable on average after $\iN \geq 4.17, 4.07, 4.01, 3.99$ for $\M = 1,2,3,4$ observations, revealing that identifiability (condition A\ref{condit} in Theorem \ref{theoId}) depends much more on the variability of the sources than on the specific mixing weights. This is confirmed in Figure \ref{fig: id_hist_mlm}, which shows the corresponding histograms and cumulative distribution functions.
The histograms and cumulative distribution functions for $\M= 1$ and $\M = 4$ differ only slightly and for $\M=1$ they look almost the same as in Figure \ref{fig: id_hist}, although in Figure \ref{fig: id_hist} $\wei$ is fixed, whereas in Figure \ref{fig: id_hist_mlm} it is random.
For the sufficient identifiability condition from Theorem \ref{theoAlgo} to hold we find an average value of $\iN \geq 15.13$ observations. Note that this condition depends on the sources only and, thus, is the same for all $\M$.

\subsection{Arbitrary Mixing Weights}

\begin{figure}[h!]
\centering
  \includegraphics[width=\linewidth]{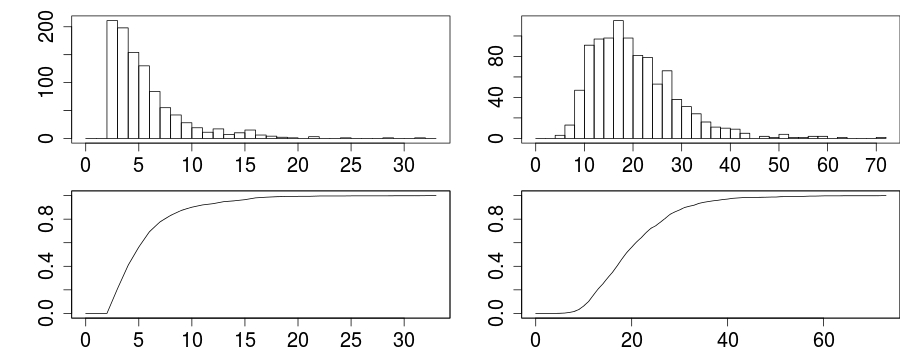}
  \caption{Top row: histograms of number of observations until $(\wei, \Sou)$ is identifiable (left) and until the sufficient identifiability condition from Theorem \ref{theoAlgo} is fulfilled (right),
bottom row: corresponding empirical cumulative distribution function,  
   with $\m = 3$, $\Al = \{-2, 1\}$, $\wei$ uniformly distributed on $\{ \wei \in [-10, 10]^\m: \wei_1 < \ldots < \wei_\m \}$ and $\Sou_{\im \iN}$ i.i.d. on $\Al$.}
  \label{fig: id_hist_aw}
\end{figure}

Finally, we consider arbitrary mixing weights in $\mathcal{A}_0$. Therefore, for each run, we draw mixing weights ($\M = 1$, $\m = 3$) independently from the uniform distribution on $\{ \wei \in [-10, 10]^\m: \wei_1 < \ldots < \wei_\m \}$ and for the sources we consider a Bernoulli model as in Section \ref{subsec:simB}.

We find that $(\wei,\Sou)$ is identifiable on average after $\iN \geq 6.09$ observations. Confirming, that identifiability, i.e., condition A\ref{condit} in Theorem \ref{theoId},
is achieved slower, when we allow for larger sets of possible mixing weights. 
For the sufficient identifiability condition from Theorem \ref{theoAlgoAbw} to hold we find an average value of $\iN \geq 20.82$ observations.  
Figure \ref{fig: id_hist_aw} shows the corresponding histograms and cumulative distribution functions.

In summary, our simulations show that the number of observations needed for our simple sufficient identifiability condition A\ref{assusuff} to hold is relatively close to the actual number of observations until $(\wei, \Sou)$ is identifiable and thus, serves as a good benchmark criterion for identifiability.
This can be used as a simple proxy for validating the applicability of any recovery procedure in practice.

\section{Conclusions}\label{sec:conc}

In this paper we have established identifiability criteria for single linear mixtures of finite alphabet sources as well as its matrix analogue. 
We gave not only sufficient but also necessary criteria for identifiability. Our work reveals the identification problem as a combinatorial problem utilizing the one to one correspondence between the mixture values and the mixing weights.
We generalized the method of Diamantaras and Chassioti \cite{diamantras2} to an arbitrary finite alphabet in order to derive a simple sufficient identifiability criterion. The proof uses the specific hierarchical structure of possible mixture values leading to successive identification of the weights.
Thus, our results characterize and extend the range of settings under which recovery algorithms (for statistical data) are applicable. 

Notably, we showed that our identifiability conditions extend to unknown number of sources $\m$. This lays the foundation to design algorithms to recover the number of active sources from a mixture and sketches a road map to pursue this in future research.

Finally, we showed that the probability of identifiability converges exponentially fast to $1$ when the underlying sources come from a discrete Markov process.
This provides a useful and simple tool to pre-determine the required number of observations in order to guarantee identifiability at a given probability.  

The derived sufficient identifiability conditions were briefly investigated in a simulation study and the required sample size for their validity  was found to be quite close to the minimal sample size for identifiability.

This work is intended to give a solid theoretical background for a model that is used in a variety of applications in digital communications, but also in bioinformatics.

\appendices
\section{Proofs}
\subsection{Proof of Theorem \ref{theoId}} \label{app:A}

\begin{IEEEproof}\quad\\
For $\sigma\in S_\m^\n$ we define $\mix^\sigma = (\mix_{\sigma(1)},\ldots, \mix_{\sigma(\m)})$.

``$\Leftarrow$ ''\\
By assumption A\ref{asstheoID} $\mix^{\rho} = \wei E $, i.e., $ \mix^{\rho} E^{-1} = \wei$
and, consequently,
\begin{align*}
\begin{aligned}
 \mix^{\rho} E^{-1} \in \mathcal{A}
\text{ and }
\{\mix_1,...,\mix_\n\} \in \big\{( \mix^{\rho} E^{-1}) \al: \al \in\Al^\m\big\},
\end{aligned}
\end{align*}
which, by assumption A\ref{condit}, is not fulfilled for any other $\sigma \in S_\m^\n$. Thus, $\wei$ is uniquely determined. Moreover, as $ASB(\wei) > 0$, $\Sou$ is uniquely determined as well.\\
``$\Rightarrow$ ''\\
Assume A\ref{condit} does not hold, i.e., there exists $\sigma \neq \rho \in S_\m^\n$ such that $\tilde{\wei} \ZuWeis  \mix^{\sigma} E^{-1}$ fulfills
\begin{align*}
\begin{aligned}
\tilde{\wei} \in \mathcal{A}
\text{ and }
\{\mix_1,...,\mix_\n\} \in \big\{\tilde{\wei} \al: \al\in\Al^\m\big\}.
\end{aligned}
\end{align*}
As we assume all observations to be pairwise different, $\tilde{\wei} \neq \wei$ and $\tilde{\wei}$ with the corresponding $\tilde{\Sou}$ lead to the same observations $\mix_1,\ldots,\mix_\n$.
Therefore, $(\wei,\Sou)$ is not identifiable.

\end{IEEEproof}

\subsection{Proof of Lemma \ref{AA}} \label{app:B}

\begin{IEEEproof}
Obviously, A\ref{assusuff} arises from A\ref{asstheoID} when we choose the matrix $E$ in Theorem \ref{theoId} as in (\ref{faMatrix}). Hence, A\ref{assusuff} implies A\ref{asstheoID} if the matrix in (\ref{faMatrix}) is invertible. (\ref{faMatrix}) can be written as 
\begin{align*}
\begin{pmatrix}
\al_1 & \ldots & \al_1 \\
& \vdots & \\
\al_1 & \ldots & \al_1 \\
\end{pmatrix}
-(\al_1-\al_2)\begin{pmatrix}
1  & \ldots & 0 \\
& \vdots & \\
0 & \ldots & 1 \\
\end{pmatrix}
\end{align*}
and consequently, the matrix in (\ref{faMatrix}) has zero determinant if and only if $\al_1-\al_2$ is an eigenvalue of 
\begin{align*}
\begin{pmatrix}
\al_1 & \ldots & \al_1 \\
& \vdots & \\
\al_1 & \ldots & \al_1 \\
\end{pmatrix},
\end{align*}
i.e., $\al_1-\al_2 = 0$ or $\al_1 - \al_2 = \m\cdot \al_1 $. As $\al_1 < \al_2$ the assertion follows.
\end{IEEEproof}

\subsection{Proof of Lemma \ref{lemmaD}} \label{app:Bb}

\begin{IEEEproof}
For $r=\m$ the assertion is obvious. So let $r< \m$. Then for $i\in \{1\ldots,k^{r}\}$ we have that
\begin{align*}
&d_i = \sum_{j=1}^{r}\wei_j B_{ji}   + \sum_{j=r+1}^{\m}\wei_j \al_1 =\\
&\sum_{j=1}^{r} \wei_j B_{ji} + \al_1(1-\sum_{j=1}^{r}\wei_j),
\end{align*}
where $B_{ji}$ are the entries of the matrix $B$ in (\ref{Bmatrix}).
\end{IEEEproof}

\subsection{Proof of Lemma \ref{lemmaCR}} \label{app:C}

\begin{IEEEproof}
By definition $\mix_{\rho(r)} = d_{k^{r-1}+1}$. Therefore, $\mix_{\rho(r)} \geq \min(\{\mix_1,\ldots,\mix_\n\} \backslash \{d_1,\ldots, d_{k^{r-1}}\})$ and for $i\in \{k^{r-1}+1,\ldots,k^\m\}$
\begin{align*}
d_i - \mix_{\rho(r)} =\sum_{l=1}^{{r-1}} \wei_l(B_{li} -\al_1) &+ \wei_{r}(B_{ri}-\al_2)\\
& + \sum_{l=r+1}^{\m} \wei_l(B_{li}-\al_1).
\end{align*}
If $B_{ri}\in\{\al_2,\ldots,\al_k\}$, then obviously $d_i - \mix_{\rho(r)} \geq 0$.
If $B_{ri}=\al_1$, then by definition of $B$ there exists an $s\in\{r+1,\ldots,\m\}$ such that $B_{si}\in\{\al_2,\ldots,\al_k\}$, and therefore,
\begin{align*}
d_i - \mix_{\rho(r)} &= \sum_{l=1}^{r-1} \wei_l (B_{li} -\al_1) + \wei_{r} (\al_1-\al_2)\\
&+ \wei_s \underbrace{(B_{si}-\al_1)}_{\geq (\al_2-\al_1)} + \sum_{l=r+1, l\neq s}^{\m} \wei_l (B_{li}-\al_1)\\
&\geq \sum_{l=1}^{r-1} \wei_l (B_{li} -\al_1) + \underbrace{(\wei_s-\wei_{r})}_{>0} \underbrace{(\al_2-\al_1)}_{>0}\\
&+ \sum_{l=r+1, l\neq s}^{\m} \wei_l(B_{li}-\al_1) \geq 0.
\end{align*}
Consequently, $\mix_{\rho(r)} \leq \min(\{\mix_1,\ldots,\mix_\n\} \backslash \{d_1,\ldots, d_{k^{r-1}}\})$.
\end{IEEEproof}

\subsection{Proof of Theorem \ref{theoexp}} \label{app:E}

\begin{IEEEproof}
Let $\fT^r$ be as in (\ref{Tr}) and let $p_0$ be the initial distribution of $(\Sou_{1 \iN},\ldots, \Sou_{\m\iN})^\top_\iN$.
Define the stopped process
\begin{align*}
\tilde{\Sou}_\iN^r \ZuWeis \begin{cases}
(\Sou_{1 \iN},\ldots, \Sou_{\m \iN})^\top & \text{ if } \iN < \fT^r\\
\sAl^r & \text{otherwise,}
\end{cases}
\end{align*}
for $r= 1, \ldots,\m$, which is a Markov process as well (see e.g., \cite[Proposition 4.11.1.]{kolokol}). It is obvious that for the Markov process $(\tilde{\Sou}_\iN^r)_\iN$ the state $\sAl^r$ is absorbing and all other states are transient. Moreover, when we reorder the states in $\Al^\m$ such that $\sAl^r$ is the first state, the transition matrix of $(\tilde{\Sou}_\iN^r)_\iN$ is given by
\begin{align*}
\tilde{P}_r= \begin{pmatrix}
1&0&\ldots &0\\
p_{2r}& &  & \\
& & Q_r & \\
p_{\m r}&  & &  \end{pmatrix}.
\end{align*}
The distribution of $\fT^r$ is a discrete phase type distribution (see e.g., \cite[Section 2.2.]{neuts}), i.e.,
\begin{align}\label{TQ}
\Pp(\fT^r > \n) =  p_0 Q_r^\n\mathbf{1} \leq \norm{Q_r^\n \mathbf{1}}_\infty.
\end{align}

As $P^\Nn > 0$ 
\begin{align*}
\tilde{P}_r^\Nn = \begin{pmatrix}
1&0&\ldots &0\\
s_{2r}&&&\\
\vdots&&Q^\Nn& \\
s_{\m r}&&& \end{pmatrix}
\end{align*}
with $s_{2r},\ldots,s_{\m r}>0$ for $r = 1, \ldots,\m$. Consequently, all row sums of $Q_r^\Nn$ are smaller than $1$, i.e.,
\begin{align}\label{cN}
c_r \ZuWeis \norm{Q_r^\Nn \mathbf{1}}_\infty <1
\end{align}
and hence $c = \max_{1 \leq r \leq \m} c_r < 1$.

Next, we show by induction that $\norm{Q_r^\n \mathbf{1}}_\infty \leq c_r^{\lfloor{n/\Nn}\rfloor}$ for all $\n \geq \Nn$.
For $\n = \Nn $ this holds by definition. So assume that $\norm{Q_r^l\mathbf{1}}_\infty \leq c_r^{\lfloor {l/ \Nn}\rfloor}$ for all $\Nn \leq l \leq \n$ and define $A =(a_{ij})_{ij}\ZuWeis Q_r^\n$, i.e.,
\begin{align*}
\max_i \sum_{j}a_{ij}\leq c_r^{\lfloor{\n / \Nn }\rfloor}.
\end{align*}
If $\lfloor \frac{\n}{\Nn }\rfloor=\lfloor \frac{\n+1}{\Nn }\rfloor$, then
\begin{align*}
&\norm{Q_r^{\n+1}\mathbf{1}}_\infty = \norm{ AQ_r\mathbf{1} }_\infty = \max_{i} \sum_j \sum_k a_{i k}q_{k j}\\
= & \max_i \sum_k a_{ik} \sum_j q_{kj} \leq c_r^{\lfloor \frac{\n}{\Nn}\rfloor} = c_r^{\lfloor \frac{\n+1}{\Nn}\rfloor},
\end{align*}
as $\max_i \sum_k a_{ik}\leq c_r^{\lfloor \frac{\n}{\Nn}\rfloor}$ and $\sum_j q_{kj} \leq 1$.

If $\lfloor \frac{\n}{\Nn }\rfloor\not=\lfloor \frac{\n+1}{\Nn}\rfloor$, then $\lfloor \frac{\n}{\Nn}\rfloor +1 =\lfloor \frac{\n+1}{\Nn}\rfloor$ and
$\n+1= \Nn \left\lfloor\frac{\n}{\Nn}\right\rfloor + \Nn$,
with $\Nn \lfloor \frac{\n}{\Nn}\rfloor \WeisZu l \in \{\Nn,\ldots,\n\}$. Therefore,
\begin{align*}
\norm{ Q_r^{\n+1}\mathbf{1}}_\infty = \norm{ Q_r^{l}Q_r^\Nn \mathbf{1}}_\infty \leq c_r^{\lfloor \frac{l}{\Nn}\rfloor+1} = c_r^{\lfloor \frac{\n+1}{\Nn}\rfloor}.
\end{align*} 
With (\ref{TQ}) and (\ref{poi}) it follows that
\begin{align*}
1-\Pp((\wei,\Sou)\text{ is identifiable})  \leq \m c^{\lfloor \frac{\n}{\Nn}\rfloor}
\end{align*}
and as
\begin{align*}
\frac{\m c^{\lfloor \frac{\n}{\Nn}\rfloor}}{c^{\frac{\n}{\Nn}}}\leq \frac{\m}{c}<\infty
\end{align*}
the assertion follows.
\end{IEEEproof}

\subsection{Proof of Theorem \ref{theoAlgoAbw}} \label{sec: apptheAlgoAbw}
\begin{IEEEproof}
Assume that $\frac{\al_2 - \al_1}{\al_k - \al_{k-1}} \leq 1 $. Otherwise, we can multiply all observations by $-1$, such that the new alphabet becomes $-\al_k < \ldots < -\al_1$, which then fulfills $\frac{\al_2 - \al_1}{\al_k - \al_{k-1}} \leq 1 $. 
Further, note that $ASB(\wei) > 0$ implies that $\wei_\im \neq 0$ for all $\im= 1,\ldots, \m$.
Let $\fG \ZuWeis \{\mix_{1}, \ldots, \mix_{\n}\}$ be the set of the pairwise different observations.
(\ref{assusuffabw}) implies that there exist $i_0,\ldots,i_{\m}, j_0, \ldots, j_\m \in \{1,\ldots,\n\}$ such that $\mix_{i_{r}} =\wei \sAl^r$ and  $\mix_{j_{r}} =\wei \sAlv^r$ for $r=0,\ldots,\m$.

First, note that 
\begin{align*}
\min \fG = \mix_{i_0} = \al_k \sum_{\substack{\im = 1 \\ \wei_\im < 0 }}^\m \wei_\im + \al_1 \sum_{\substack{\im = 1 \\ \wei_\im > 0} }^\m \wei_\im, \\
\max \fG = \mix_{j_0} = \al_1 \sum_{\substack{\im = 1 \\ \wei_\im < 0}}^\m \wei_\im + \al_k \sum_{\substack{\im = 1 \\ \wei_\im > 0}}^\m \wei_\im
\end{align*}
and thus
\begin{align*}
\fo_+ \ZuWeis \sum_{\substack{\im = 1 \\ \wei_\im > 0 }}^\m \wei_\im &= \frac{\al_k \max \fG - \al_1 \min \fG }{\al_k^2 - \al_1^2},\\
\fo_- \ZuWeis \sum_{\substack{\im = 1 \\ \wei_\im < 0 }}^\m \wei_\im &= \frac{\al_k \min \fG - \al_1 \max \fG}{\al_k^2 - \al_1^2}.
\end{align*}
If $\fo_- = 0$,  all weights are positive and, as $\fo_+$ is identified and thus w.l.o.g equal to one, Theorem \ref{theoAlgo} applies. Thus, assume that $\fo_- < 0$ and define $\fG_0 \ZuWeis \fG \backslash \{\min\fG, \max \fG\} $ and
\begin{align*}
\tilde{\m}_0 &\ZuWeis  
\max \{\im = 1,\ldots,\m \text{ s.t. } \wei_\im < 0 \},\\
\tilde{\m}_0^+ &\ZuWeis \tilde{\m}_0 + 1,
\end{align*}
i.e., $\wei_1 < \ldots < \wei_{\tilde{\m}_0} < 0 < \wei_{\tilde{\m}_0^+} <\ldots <\wei_\m $. 
Second, note that analog to (\ref{remarkC1})
\begin{align*}
\min \fG_0 = \min(\mix_{i_{\tilde{\m}_0^+}},\mix_{j_{\tilde{\m}_0}} ),\quad
\max \fG_0 = \max(\mix_{j_{\tilde{\m}_0^+}}, \mix_{i_{\tilde{\m}_0}})
\end{align*}
and thus
\begin{align*}
\frac{\min \fG_0  - \al_k \fo_- - \al_1 \fo_+}{\al_k - \al_{k-1}} &= \min\left(\frac{\al_2 - \al_1}{\al_k - \al_{k-1}}\wei_{\tilde{\m}_0^+},-\wei_{\tilde{\m}_0}\right)\\
\frac{ \al_k \fo_+ + \al_1 \fo_- - \max \fG_0}{\al_2 - \al_1} &= \min\left(-\wei_{\tilde{\m}_0}, \frac{\al_k - \al_{k-1}}{\al_2 - \al_1}\wei_{\tilde{\m}_0^+}\right).
\end{align*}
Hence, if $\frac{\min \fG_0  - \al_k \fo_- - \al_1 \fo_+}{\al_k - \al_{k-1}}  < \frac{ \al_k \fo_+ + \al_1 \fo_- - \max \fG_0}{\al_2 - \al_1}  $ we find that
\begin{align*}
\wei_{\tilde{\m}_0^+} = \frac{\min \fG_0  - \al_k \fo_- - \al_1 \fo_+}{\al_2 - \al_{1}} 
\end{align*}
and if $\frac{\min \fG_0  - \al_k \fo_- - \al_1 \fo_+}{\al_k - \al_{k-1}}  = \frac{ \al_k \fo_+ + \al_1 \fo_- - \max \fG_0}{\al_2 - \al_1}  $ that
\begin{align*}
\wei_{\tilde{\m}_0} = \frac{\max \fG_0 -  \al_k \fo_+ - \al_1 \fo_- }{\al_2 - \al_1}.
\end{align*}
Thus, we have identified the first weight, namely
\begin{align*}
\wei_1^{\star} \ZuWeis \begin{cases}
\wei_{\tilde{\m}_0^+} & \text{ if } \frac{\min \fG_1  - \al_k \fo_- - \al_1 \fo_+}{ \al_k \fo_+ + \al_1 \fo_- - \max \fG_1}  < \frac {\al_k - \al_{k-1}}{\al_2 - \al_1}  \\
\wei_{\tilde{\m}_0} & \text{ otherwise}.
\end{cases}
\end{align*}
Now assume that we have identified $l$ different weights, $\wei_1^{\star},\ldots,\wei_l^{\star}$. If $\fo_- = \sum_{\substack{\im = 1 \\ \wei_\im^{\star} < 0}}^l \wei_\im^{\star}$, all the remaining weights are positive and Theorem \ref{theoAlgo} applies. Thus assume that $\fo_- < \sum_{\substack{\im = 1 \\ \wei_\im^{\star} < 0}}^l \wei_\im^{\star}$
and define $\fG_l \ZuWeis \fG_{l-1} \backslash \fR_{l-1}$, with
\begin{align*}
\fR_{l-1} \ZuWeis \bigcup_{\substack{\al^{\prime}, \al^{\prime \prime} \in \Al \\ \al \in \Al^l }}\Big\{ \al^{\prime}(\fo_- - \sum_{\substack{\im = 1 \\ \wei_\im^{\star}<0 }}^l \wei_\im^{\star} )
&+ \al^{\prime \prime} (\fo_+  - \sum_{\substack{\im = 1 \\ \wei_\im^{\star}>0 }}^l \wei_\im^{\star} )\\
& +(\wei_1^{\star},...,\wei_l^{\star}) \al  \Big\}
\end{align*}
and
\begin{align*}
\tilde{\m}_l &\ZuWeis 
\max \{i = 1,...,\m \text{ s.t. } \wei_\im < 0 \text{ and } \wei_\im \not\in \{\wei_1^{\star},...,\wei_l^{\star}\} \} ,\\
\tilde{\m}_l^+ &\ZuWeis 
\min \{\im = 1,...,\m \text{ s.t. } \wei_\im > \wei_{\tilde{\m}_l} \text{ and } \wei_\im \not\in \{\wei_1^{\star},...,\wei_l^{\star}\} \}.
\end{align*}
Note that analog to Lemma \ref{lemmaCR}
\begin{align*}
\min \fG_l = \min(\mix_{i_{\tilde{\m}_l^+}},\mix_{j_{\tilde{\m}_l}} ), \quad
\max \fG_l = \max(\mix_{j_{\tilde{\m}_l^+}}, \mix_{i_{\tilde{\m}_l}})
\end{align*}
and thus
\begin{align*}
\frac{\min \fG_l  - \al_k \fo_- - \al_1 \fo_+}{\al_k - \al_{k-1}} &= \min\left(\frac{\al_2 - \al_1}{\al_k - \al_{k-1}}\wei_{\tilde{\m}_l^+},-\wei_{\tilde{\m}_l}\right)\\
\frac{ \al_k \fo_+ + \al_1 \fo_- - \max \fG_l}{\al_2 - \al_1} &= \min\left(-\wei_{\tilde{\m}_l}, \frac{\al_k - \al_{k-1}}{\al_2 - \al_1}\wei_{\tilde{\m}_l^+}\right).
\end{align*}
Hence, if $\frac{\min \fG_l  - \al_k \fo_- - \al_1 \fo_+}{\al_k - \al_{k-1}}  < \frac{ \al_k \fo_+ + \al_1 \fo_- - \max \fG_l}{\al_2 - \al_1}  $ we find that
\begin{align*}
\wei_{\tilde{\m}_l^+} = \frac{\min \fG_l  - \al_k \fo_- - \al_1 \fo_+}{\al_2 - \al_{1}} 
\end{align*}
and if $\frac{\min \fG_l  - \al_k \fo_- - \al_1 \fo_+}{\al_k - \al_{k-1}}  = \frac{ \al_k \fo_+ + \al_1 \fo_- - \max \fG_l}{\al_2 - \al_1}  $ that
\begin{align*}
\wei_{\tilde{\m}_l} = \frac{\max \fG_l -  \al_k \fo_+ - \al_1 \fo_- }{\al_2 - \al_1}.
\end{align*}
Thus, we have identified the $(l+1)$-th weight as
\begin{align*}
\wei_{l+1}^{\star} \ZuWeis \begin{cases}
\wei_{\tilde{\m}_l^+} & \text{ if } \frac{\min \fG_l  - \al_k \fo_- - \al_1 \fo_+}{ \al_k \fo_+ + \al_1 \fo_- - \max \fG_l}  < \frac {\al_k - \al_{k-1}}{\al_2 - \al_1}  \\
\wei_{\tilde{\m}_l} & \text{ otherwise}.
\end{cases}
\end{align*}
By induction, we can identify all weights and thus, by $ASB(\wei)>0$ the assertion follows.
\end{IEEEproof}

\section*{Acknowledgment}
The authors acknowledge support of DFG CRC 803, 755, RTG 2088, and FOR 916. Helpful comments of an editor, two referees, C. Holmes, P. Rigollet, and H. Sieling are gratefully acknowledged.

\bibliographystyle{IEEEtran}

\end{document}